\documentstyle[preprint,prl,aps,epsf]{revtex}

\begin{document}
\preprint{LBNL-41764, May 1998}
\title{Spatial Variation of Nuclear Structure Functions \\
and Heavy Quark Production}

\author{V. Emel'yanov$^1$, A. Khodinov$^1$, S. R. Klein$^2$ and
R. Vogt$^{2,3}$}
 
\address{{
$^1$Moscow State Engineering Physics Institute (Technical
University), Kashirskoe ave. 31, Moscow, 115409, Russia\break 
$^2$Nuclear Science Division, Lawrence Berkeley National Laboratory, 
Berkeley, CA
94720, USA\break 
$^3$Physics Department, University of California, Davis, CA 95616, USA}\break}
 
\vskip .25 in
\maketitle
\begin{abstract}

We explore how nuclear modifications to the free nucleon structure functions
(also known as shadowing) affect heavy quark production in collisions at
different impact parameters.  We assume that the nuclear modifications
arise from a density dependent effect such as gluon recombination and are thus
proportional to the local density. We calculate the dependence of charm and
bottom quark production on impact parameter and
show that density dependent modifications can lead to significant reductions 
in the heavy quark production cross sections in central relative to peripheral
interactions.

\vskip .3 in
\centerline{(Submitted to Physical Review Letters.)}

\end{abstract}

Experiments \cite{Arn} have shown that the proton and neutron
structure functions are modified by a nuclear environment.  For
momentum fractions $x< 0.1$ and $0.3<x<0.7$, a depletion is observed
in the nuclear parton distributions.  The low $x$, or shadowing,
region and the larger $x$, or EMC region, is bridged by an enhancement
known as antishadowing for $0.1 < x < 0.3$.  Recently, the entire
characteristic modification as a function of $x$ has been referred to
as shadowing.  Many theoretical explanations of the nuclear effect
have been proposed.  In most, the degree of modification depends on
the local nuclear density.  For example, if the modification is due to
gluon recombination \cite{hot}, then the degree of modification should
depend directly on the local gluon density and hence on the spatial
position of the interaction within the nucleus.  Nuclear binding and
rescaling models also predict that the structure function depends on
the local density \cite{close}.

Most measurements of structure functions have used lepton or neutrino
beams.  Typically these experiments are insensitive to the location of
the interaction witin the nucleus so that the resulting structure
functions are averaged over the entire nucleus.  The Fermilab
E745 collaboration studied $\nu N$ interactions %with dark tracks 
in a bubble chamber where the dark tracks indicated that the interaction
occurred deep within the nucleon.  They found that structure function
does vary spatially, but had little direct sensitivity to the impact
parameter\cite{e745}.

This letter discusses the effects of spatially-dependent nuclear
parton distributions
on charm and bottom production cross sections as a function of impact
parameter.  Assuming that shadowing is proportional to
the local nucleon density, we study the dependence of heavy quark production
on the nucleon structure functions and the shadowing parameterization.

We show that the heavy quark production cross section changes
significantly when this spatial dependence is considered.  The
variation with impact parameter is especially important because
relatively peripheral collisions are often used as a baseline in
searches for new phenomena in more central collisions\cite{na50}.  
Large impact parameter collisions tend to probe the nuclear surface
where shadowing is rather weak
while central collisions are also sensitive to structure functions at
the nuclear core where nuclear modifications can be large.

The impact parameter of an event can be determined from the transverse
energy produced in the event.  The spatial dependence of shadowing
on transverse energy production has already been
considered\cite{Eskola2}. In Au+Au collisions at $\sqrt{s_{NN}} = 200$ GeV,
the center of mass energy
per nucleon pair, a typical calorimeter can
measure impact parameters as large as $b=1.8 R_A$ where $R_A$ is the
nuclear radius\cite{us}.

Heavy quark production in heavy ion collisions is dominated by gluon
fusion\cite{Alde,Appel}.  At leading order (LO), charm and bottom
quarks are produced in two basic processes: $q\overline q\rightarrow
Q\overline Q$ and $gg\rightarrow Q\overline Q$ with $Q = c$ and $b$.  The LO
cross section for a nucleus A with momentum $P_A$ colliding with
nucleus B with momentum $P_B$ at an impact parameter $b$ is then
\begin{eqnarray}
\lefteqn{ E_Q E_{\overline Q} \frac{d\sigma_{AB}}{d^3p_Q d^3p_{\overline Q} 
d^2b d^2r} = } \label{sigma} \\ & & \sum_{i,j}\int \,dz \,dz'
\,dx_1\, dx_2 F_i^A(x_1,Q^2,\vec{r},z) F_j^B(x_2,Q^2,\vec{b} - \vec{r},z')
E_Q E_{\overline Q} \frac{d\widehat{\sigma}_{ij}
(x_1P_A,x_2P_B,m_Q,Q^2)}{d^3p_Q d^3p_{\overline Q}} \, \,  .
\nonumber 
\end{eqnarray}
where $Q$ represents the produced heavy quark.  Here $i$ and $j$ are
the interacting partons in the nucleus and the functions $F^A_i$ and
$F^B_i$ are the number densities of gluons, light quarks and
antiquarks in each nucleus, evaluated at momentum fraction $x$,
momentum scale $Q^2$, and location $\vec{r}$, $z$ in
the two nuclei.  The short-distance cross section,
$\widehat{\sigma}_{ij}$, is calculable as a perturbation series in the
strong coupling constant $\alpha_s(Q^2)$, see Ref. \cite{Ellis}.

These leading order calculations underestimate the measured charm
production cross sections by a constant, usually called the $K$
factor.  A similar theoretical factor, $K_{\rm th}$, is given by the ratio
of the next-to-leading-order (NLO) cross section to the leading order
result.  The $K$ factor is relatively independent of the quark $p_T$,
pair invariant mass and rapidity distribution, and the parton
density\cite{HPC}.  However, it can vary significantly with quark mass
and beam energy.  

Since this calculation is at leading order, 
we use the GRV 94 LO \cite{GRV94}
nucleon parton distributions, evaluated with $m_c=1.3$ GeV, $m_b=4.75$ GeV and 
$Q=m_T$ where $m_T^2 = p_T^2 + m_Q^2$. 
We also used the GRV 94 HO \cite{GRV94} with the same parameters
and the MRS G distributions
\cite{MSRG}, evaluated with the same parameters for bottom and for charm,
$m_c=1.2$ GeV and $Q = 2m_T$.  The mass and scale parameters used with each
set were chosen for their agreement with the $Q \overline Q$ total cross
section data.

We assume that the parton densities $F_i^A(x,Q^2,\vec{r},z)$ can be
represented as the product of $x$ and $Q^2$ independent nuclear
density distributions, position and $A$ independent nucleon parton
densities, and a shadowing function that contains the modification of
the nuclear structure functions.
\begin{eqnarray}
F_i^A(x,Q^2,\vec{r},z) & = & \rho_A(s) S^i(A,x,Q^2,\vec{r},z) 
f_i^p(x,Q^2) \\
F_j^B(x,Q^2,\vec{b} - \vec{r},z') & = & \rho_B(s') S^j(B,x,Q^2,\vec{b} - 
\vec{r},z') f_j^p(x,Q^2) \nonumber \,\, ,
\end{eqnarray}
where $f^p(x,Q^2)$ are the nucleon parton densities, $s = \sqrt{r^2 +
z^2}$ and $s'=\sqrt{|\vec{b}-\vec{r}|^2 +z'^{\, 2}}$.  In the absence
of nuclear modifications of the structure functions (no shadowing),
$S^i(A,x,Q^2,\vec{r},z)\equiv1$.  The nuclear density is given by a
Woods-Saxon distribution
\begin{equation}
\rho_A(s)= \rho_0 {1 + \omega(s/R_A)^2 \over 1 + \exp[(s-R_A)/d]} 
\label{density}
\end{equation}
where electron scattering data from \cite{Vvv} is used to fix 
the parameters $R_A$, $d$, $\omega$ and $\rho_0$.

The shadowing effect is studied with two parameterizations previously used to
estimate the effect on heavy quark production without including
spatial dependence\cite{GMRV}. The first, $S_1(A,x)$ is based on fits
to recent nuclear deep inelastic scattering data\cite{EQC}.  It treats
the quark, gluon and antiquark functions equally without $Q^2$
evolution. The second,
$S_2^i(A,x,Q^2)$ has separate modifications for the valence quarks, sea quarks
and gluons and includes $Q^2$ evolution\cite{KJE}.

To include the spatial dependence of shadowing, we assume
that these modifications are proportional to
the undisturbed local nuclear rest density, Eq.~(\ref{density}), 
\begin{eqnarray}
S^i_{\rm WS} = 
S^i(A,x,Q^2,\vec{r},z) =  1 + N_{\rm WS}
[S^i(A,x,Q^2) - 1] \frac{\rho(s)}{\rho_0} \label{wsparam} 
\end{eqnarray}
where $N_{\rm WS}$%=1.317$ 
is a normalization constant chosen such that
$(1/A) \int d^3 s \rho(s) S^i_{\rm WS} = S^i$. At large radii, $s \gg
R_A$, medium modifications weaken so that in very peripheral
interactions, nucleons behave act as if they are in free space.  At
the center of the nucleus, the modifications are larger than the
average value found in lepton scattering experiments.  A second
parameterization, $S_{\rm R}^i$, based on the thickness of a spherical
nucleus at the collision point\cite{us}, leads to a slightly larger
modification in the nuclear core.  Because it assumes a sharp nuclear
surface, for $s \geq R_A$, $S_{\rm R}^i = 1$, enhancing the shadowing
at the center over that found with Eq.~(\ref{wsparam}).  
%The corresponding normalization constant is $N_R = 1.449$.  
Thus for surface nucleons, Eq.~(\ref{wsparam}) predicts modifications half as
strong as at the core while no further modifications are predicted
with $S_{\rm R}^i$.  Other calculations assuming distinct shadowing
effects in different nuclear orbitals\cite{close} cannot be directly
compared to our results.

With these ingredients, Eq.~(\ref{sigma}) can be used to find the
charm and bottom cross sections as a function of impact parameter.
Table 1 gives the total cross sections in nucleus-nucleus collisions
integrated over impact parameter (in units of $\mu$b/nucleon pair) for
several cases.  As a baseline for comparison we give the results with
$S=1$ for GRV 94 HO and MRS G structure functions, followed by $S=1$,
$S_1$, and $S_2$ for GRV 94 LO.  The theoretical $K$ factors for each
$S=1$ set are also included.  We note that, with this normalization,
the total cross section, integrated over all impact parameters, is
unchanged when spatial dependence is included in $S^i$.

Although the spatial
parameterizations are normalized to reproduce the impact-parameter integrated
cross sections, the total cross section changes with impact parameter
in heavy ion collisions.  This effect is illustrated for the GRV 94 LO
distributions in Figs.\
1 and 2.  To emphasize the role of the spatial dependence, the results are
given relative to the $S=1$ cross section.  Figure 1
presents the charm production ratios for collisions at LHC
(Pb+Pb at $\sqrt{s_{NN}}=5.5$ TeV), RHIC (Au+Au at
$\sqrt{s_{NN}}=200$ GeV) and the SPS
(Pb+Pb at $\sqrt{s_{NN}}=17.3$ GeV) 
while Fig. 2 shows the bottom production ratios at LHC and RHIC.

Figure 1 shows that the nuclear effect increases with the collision energy. 
At RHIC and LHC, charm production occurs predominantly at momentum
fractions $x_i<0.1$, 
in the shadowing region, leading to a decrease in the cross
sections for both $S_1$ and $S_2$.  In central collisions at the LHC, the
cross section decreases by a nearly 50\% while at RHIC the
decrease is about 30\%.  In both cases, as the impact parameter increases,
somewhat stronger shadowing is observed in central collisions when the 
spatial dependence is included while for impact parameters larger than $R_A$
the shadowing decreases until the ratio begins to approach unity for $b >
2R_A$.  At the much lower SPS energy, charm production typically occurs in the
antishadowing region.  Since parameterization $S_2$ includes stronger gluon
antishadowing than $S_1$, the charm cross section is enhanced by 12\% in
central collisions when
$S_2$ is used while the nuclear effect with $S_1$ is negligible.  As the
impact parameter increases, the enhancement decreases.  Similar effects
are seen with the other sets of parton distributions studied.
 
We note that these cross sections are integrated over the final-state 
kinematics and that, under certain kinematic conditions, the shadowing
effect can be larger. 
Thus, the rapidity and $p_T$ distributions of the produced quarks
may exhibit a stronger shadowing spatial dependence, as shown in detail
at RHIC energies in Ref.~\cite{us}. {\it E.g.} when $y=0$ for both charm 
quarks, $x_1 = x_2$, providing the cleanest determination of the
magnitude of the nuclear modification.

Figure 2 shows the corresponding shadowing ratios for $b\overline b$
production at LHC and RHIC.  The heavier $b$ quark probes larger
values of $x$ and $Q^2$ than the charm quark at the same energies and
thus exhibit a different shadowing pattern.  At the lower $Q^2$ of
charm production, the scale-dependent shadowing parameterization $S_2$
produces stronger gluon shadowing than $S_1$ at LHC and RHIC.
However, for $b \overline b$ production, the $Q^2$ dependence of $S_2$
reduces the gluon shadowing with respect to $S_1$.  Figure 1(a) shows
an approximate 60\% shadowing at $b\approx 0$ with $S_2$, compared
with 25\% for $b$ quarks in Fig.~2(a).  Meanwhile calculations with
$S_1$ show little difference between charm and bottom production.
This is because at the typical $x$ values at LHC energies, $\sim 4.7
\times 10^{-4}$ and $1.7 \times 10^{-3}$ for $c$ and $b$ quarks
respectively at $y=0$ and $p_T = 0$, nuclear shadowing has been seen
to saturate \cite{E6652} so that the larger $x$ for $b$ production has
only a small effect on the $Q^2$-independent $S_1$ parmeterization.
The lower energy of RHIC, where $x \sim 0.013$ for charm and 0.048 for
bottom with $y=0$ and $p_T =0$, exhibits a larger difference between
the heavy quarks, especially after $Q^2$ evolution is considered.
Over the kinematic range of the total cross section, $b \overline b$
production results in a 2\% enhancement in central collisions due to
antishadowing for $S_2$.  A 9\% depletion is still observed for $S_1$
without $Q^2$ evolution.  Although not shown, we note that at SPS
energies, $b$ production is predominantly in the EMC region of $x$,
leading again to a depletion of bottom production relative to that
observed with no shadowing.

Our results show that using peripheral collisions as a baseline
for comparison to central collisions can lead to significant errors if the
spatial dependence of shadowing is not taken into account.  At RHIC and LHC,
the cross section is generally higher in peripheral collisions than might be
expected without any spatial dependence of shadowing.  Interestingly this
effect is reversed at the SPS.  We note however that the spatial dependence of
the shadowing is not evident until $b > 1.2 R_A$, requiring studies of
peripheral collisions to determine the strength of the effect.

Similar impact parameter based effects should be observable in other
hard processes such as J/$\psi$ and Drell-Yan
production \cite{us2}.  The Drell-Yan pairs\cite{cast}
are of special interest because they are produced by
$q\overline q$ interactions at leading order, 
in contrast to the gluon-dominated heavy quark production.

In conclusion, we have shown that introducing a very natural spatial
dependence in nuclear shadowing changes the heavy quark production rates in
heavy ion collisions, altering the expected relationship between central and 
peripheral collisions.  This alteration could lead to a misinterpretation of
the transverse energy dependence of certain quark-gluon plasma signatures
due to the relationship between transverse energy and impact parameter.

V.E. and A.K. would like to thank the LBNL Relativistic Nuclear
Collisions group for their hospitality and M. Strikhanov and V.V. Grushin for
discussions and support.  We also thank K.J. Eskola for providing the
shadowing routines and for discussions.  This work was supported in
part by the Division of Nuclear Physics of the Office of High Energy
and Nuclear Physics of the U. S. Department of Energy under Contract
Number DE-AC03-76SF0098.

\begin{figure}[h]
\setlength{\epsfxsize=0.7\textwidth}
\setlength{\epsfysize=0.7\textheight}
\centerline{\epsffile{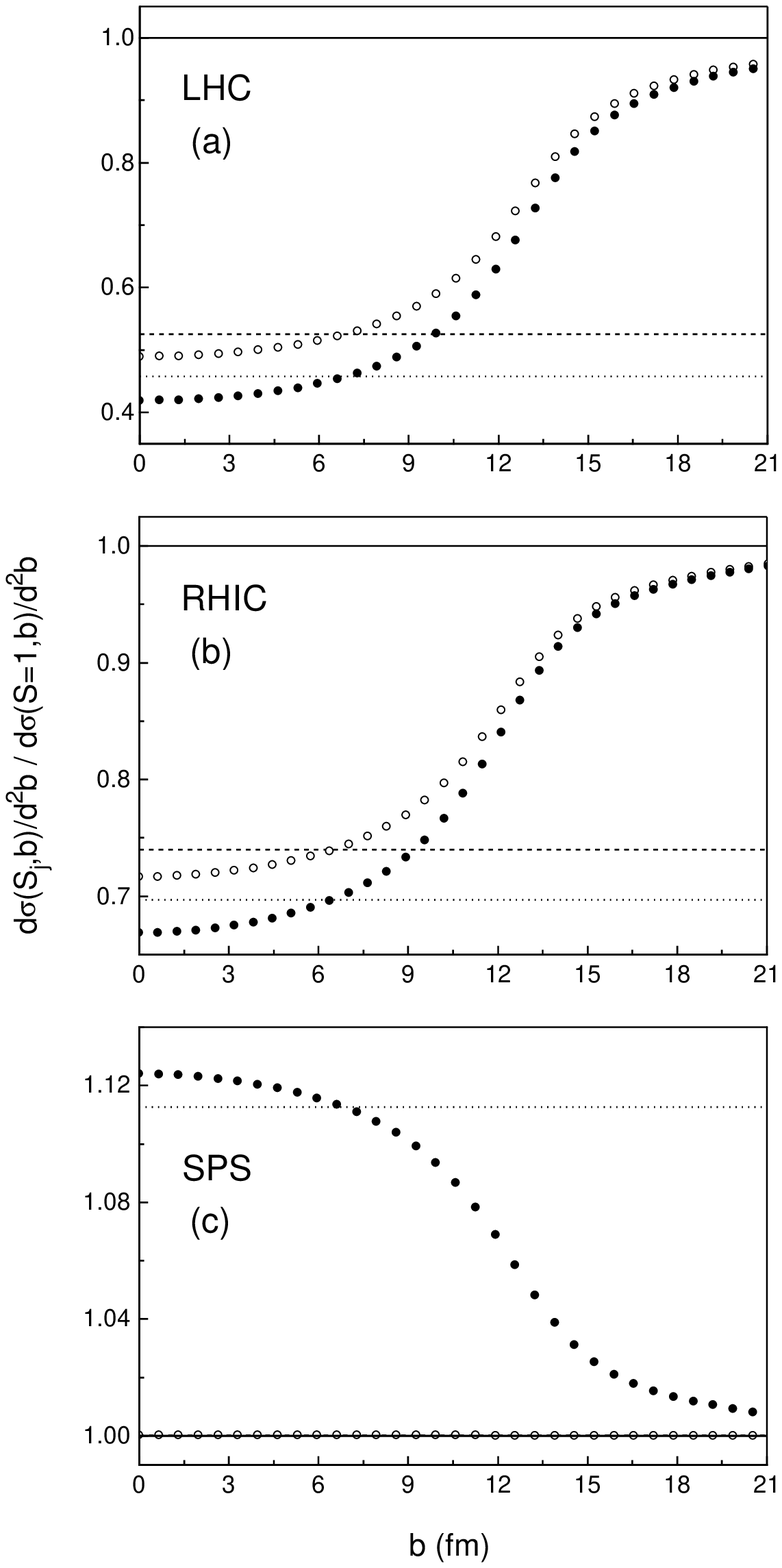}}
\caption{Charm quark production relative to production without
shadowing, $S=1$, as a function of impact parameter.  The dashed and
dotted lines show the effect with shadowing but without spatial
dependence for $S_1$ and $S_2$ respectively.  The spatial dependence
is illustrated for $S_{1\,{\rm WS}}$ (open circles) and $S_{2\,{\rm
WS}}$ (filled circles).  The results are shown for (a) Pb+Pb
collisions at the LHC with $\sqrt{s_{NN}} = 5.5$ TeV, (b) Au+Au
collisions at RHIC with $\sqrt{s_{NN}} = 200$ GeV and (c) Pb+Pb
collisions at the CERN SPS with $\sqrt{s_{NN}} = 17.3$ GeV.}
\end{figure}

\pagebreak
\begin{figure}[h]
\setlength{\epsfxsize=0.8\textwidth}
\setlength{\epsfysize=0.7\textheight}
\centerline{\epsffile{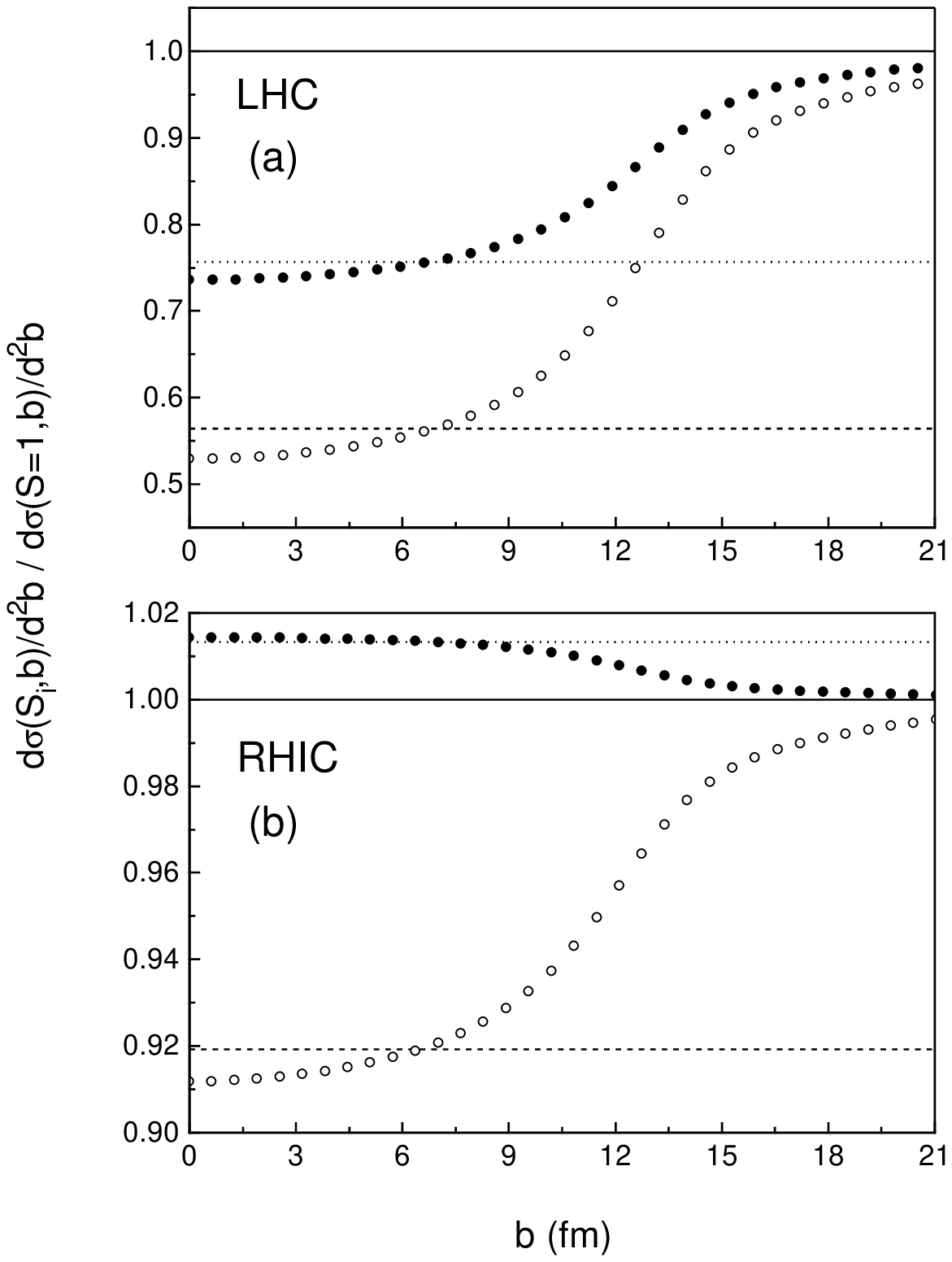}}
\caption{Bottom quark production relative to production without
shadowing, $S=1$, as a function of impact parameter.  The dashed and
dotted lines show the effect with shadowing but without spatial
dependence for $S_1$ and $S_2$ respectively.  The spatial dependence
is illustrated for $S_{1\,{\rm WS}}$ (open circles) and $S_{2\,{\rm
WS}}$ (filled circles).  The results are shown for (a) Pb+Pb
collisions at the LHC with $\sqrt{s_{NN}} = 5.5$ TeV and (b) Au+Au
collisions at RHIC with $\sqrt{s_{NN}} = 200$ GeV.}
\end{figure}

\begin{table}
\begin{tabular}{lccccccccc}
$\sqrt{s_{NN}}$ &   & \multicolumn{2}{c}{MRS G} & \multicolumn{2}{c}{GRV 94 HO}
& \multicolumn{4}{c}{GRV 94 LO} \\
(GeV) & $Q \overline Q$ & $\sigma(S=1)$ & 
$K_{\rm th}$ & $\sigma(S=1)$  & $K_{\rm th}$ &  $\sigma(S=1)$
&  $K_{\rm th}$ & $\sigma(S_1)$ & $\sigma(S_2)$ \\ 
\hline
17.3 & $c\overline c$ & 1.53 & 2.85 & 1.80 & 2.35 & 1.85 & 2.36 
& 1.85 & 2.06 \\
200  & $c\overline c$ & 138 & 2.31 & 107 & 2.35 & 174 & 2.46 
& 129 & 121 \\
200  & $b\overline b$ & 0.693 & 1.87 & 0.702 & 1.84 & 0.940 & 1.89 
& 0.866 & 0.953 \\
5500 & $c\overline c$ & 5622 & 2.10 & 2130 & 2.77 & 7440 & 2.69 
& 3910 & 3410 \\
5500 & $b\overline b$ & 93.7 & 1.78 & 85.7 & 1.71 & 212 & 1.80 
& 120 & 160 \\
\end{tabular}
\caption{Leading order $c \overline c$ and $b \overline b$ total cross
sections, in units of $\mu$b per nucleon pair, integrated over all
impact parameters, for the MRS G, GRV 94 HO and GRV 94 LO parton
distribution as well as the theoretical $K$ factors ($K_{\rm th} =
\sigma_{\rm NLO}/\sigma_{\rm LO}$) for $S=1$.  The cross sections
including shadowing are also given for the GRV 94 LO distributions.
The quark masses and the scale parameters are described in the text.}
\end{table}

\begin{references}

\bibitem{Arn} J.J. Aubert {\it et al.}, Nucl. Phys. {\bf B293} 740, (1987);
M. Arneodo, Phys. Rep. {\bf 240} 301, (1994).

\bibitem{hot} L.V. Gribov, E.M. Levin, and M.G. Ryskin, Phys. Rep. {\bf 100} 
1, (1983).

\bibitem{close} S. Kumano and F.E. Close, Phys. Rev. C {\bf 41}, 1855
(1990).

\bibitem{e745} T. Kitagaki {\it et al.}, Phys. Lett. {\bf 214}, 281
(1988).

\bibitem{na50} M.C. Abreu {\it et al.} (NA50 Collab.), Phys. Lett. {\bf B410}
(1997) 327, 337

\bibitem{Eskola2} K.J. Eskola, Z. Phys. C {\bf 51} 633 (1991).

\bibitem{us} V. Emel'yanov, A. Khodinov, S.R. Klein and R. Vogt,
Phys. Rev. {\bf C56}, 2726 (1997); V. Emel'yanov, A. Khodinov and
M. Strikhanov, Yad. Fiz. {\bf 60}, 539 (1997) [Phys. of Atomic Nuclei,
{\bf 60} 465, (1997)].
 
\bibitem{Alde} D.M. Alde {\it et al.}, Phys. Rev. Lett. {\bf 66} 133,
(1991).
 
\bibitem{Appel} J.A. Appel, Ann. Rev. Nucl. Part. Sci. {\bf 42} 367, (1992).
 
\bibitem{Ellis} R.K. Ellis, in {\it Physics at the 100 GeV Scale},
Proceedings of the 17th SLAC Summer Institute, Stanford, California,
1989, edited by E.C. Brennan (SLAC Report No. 361, Stanford, 1990).

\bibitem{HPC} P.L. McGaughey {\it et al.}, Int. J. Mod. Phys. {\bf A10}
2999, (1995).

\bibitem{GRV94} M. Gl\"{u}ck, E. Reya, and A. Vogt,
Z. Phys. {\bf C67} 433, (1995).

\bibitem{Vvv}
C.W. deJager, H. deVries, and C. deVries, Atomic Data and Nuclear Data 
Tables {\bf 14} 485, (1974).

\bibitem{MSRG} A.D.~Martin, R.G.~Roberts and W.J. Stirling,
Phys. Lett. {\bf B354} 155, (1995).

\bibitem{GMRV} S. Gavin, P.L. McGaughey, P.V. Ruuskanen and R. Vogt, Phys.
Rev. {\bf C54} 2606, (1996).

\bibitem{EQC}
K.J. Eskola, J. Qiu, and J. Czyzewski, private communication.

\bibitem{KJE}
K.J. Eskola, Nucl. Phys. {\bf B400} 240, (1993).

%\bibitem{RHIC} {\it Conceptual Design for the Relativistic Heavy Ion
%Collider}, BNL-52195, May, 1989, Brookhaven National Laboratory.
 
% \bibitem{pdflib} H. Plothow-Besch, Comp. Phys. Comm. {\bf 75} 396,
%(1993).

\bibitem{E6652} M.R. Adams {\it et al.}, Phys. Rev. Lett. {\bf 68} 
3266, (1992).

\bibitem{us2} V. Emel'yanov, A. Khodinov, S.R. Klein and R. Vogt,
in preparation.

\bibitem{cast} P. Castorina and A. Donnachie, Z. Phys. C {\bf 49},
481 (1991). 
  
\end{references}
\end{document}